\documentclass[prl,twocolumn,showpacs,amsmath,amssymb]{revtex4-1}
\usepackage{graphicx}
\usepackage{dcolumn}
\usepackage{bm}
\def\3he{$^3$He}
\def\4he{$^4$He}

\begin{document}


\title{Universal temperature dependence, flux extinction and the role of $^3$He impurities in superfluid mass transport through solid $^4$He}


\author{Ye. Vekhov}
\author{W.J. Mullin}
\author{R.B. Hallock}%
\affiliation{%
Laboratory for Low Temperature Physics, Department of Physics,\\
University of Massachusetts, Amherst, MA 01003}
\date{June 23, 2014}%


\begin{abstract}

  The mass flux, $F$, carried by  as-grown solid \4he in the range $25.6 - 26.3$ bar  rises with falling temperature and at a concentration-dependent temperature, $T_d$, the flux decreases sharply within a few mK.   We study $F$ as a function of  \3he impurity concentration, $\chi$. We find that $T_d$ is an increasing function of increasing $\chi$.  At temperatures above $T_d$ the flux has a  universal temperature dependence and the flux terminates in a narrow window near a characteristic temperature $T_h \approx$ 625 mK, which is independent of $\chi$.

\end{abstract}

\maketitle

The torsional oscillator measurements on solid \4he by Kim and Chan\cite{Kim2004a,Kim2004b}
stimulated substantial interest in the properties of solid \4he.   Changes in the stiffness\cite{Day2007} of \4he were found to have a temperature dependence very similar to the period shifts seen in the torsional oscillator work.  More
recent work by Chan's group that utilized a Vycor cell coated with epoxy  (that contained no bulk solid \4he) showed no significant period shifts\cite{Kim2012}.
It is now believed by many that the original Kim and Chan\cite{Kim2004a,Kim2004b} observations resulted from changes in the stiffness\cite{Day2007} of the bulk helium in the sample cell and the influence of this temperature-dependent stiffness on the torsional oscillator itself and not from supersolidity.

Experiments designed to create mass flow in solid \4he in confined geometries by directly squeezing the solid lattice have not been successful\cite{Greywall1977,Day2005,Day2006,Rittner2009}.  But, by the creation of chemical potential differences across bulk solid samples in contact with superfluid helium, a mass flux, $F$, has been documented\cite{Ray2008a,Ray2009b}.   For \4he with a nominal 0.3 ppm \3he content those experiments revealed the strong temperature dependence\cite{Ray2010c,Ray2011a} of $F$ at $T_d \approx 75-80$~mK, with behavior at higher temperatures that indicated the presence of a Bosonic Luttinger liquid\cite{Vekhov2012,DelMaestro2010,DelMaestro2011}.  The details of what is definitively responsible for this have not been established. The results to date are consistent with dissipative superflow along one-dimensional dislocation cores\cite{Soyler2009}, but alternate scenarios have been suggested\cite{Sasaki2008}. In the present work we report measurements of $F$ and $T_d$, as a function of \3he impurity concentration, $\chi$, in the pressure range $25.6 - 26.3$ bar and conclude that the extinction of the flux at $T_d$ is related \3he leaving the solid mixture and blocking the flux carriers. For $T > T_d$ the flux is sample-dependent, has a  universal temperature dependence and the flux terminates in a narrow window near a characteristic temperature $T_h \approx$~625 mK, which is independent of $\chi$.  Cooling through $T_d$ the flux drops sharply within a few mK.

 Since the apparatus\cite{Ray2010c,Ray2011a} used for this work has been illustrated and described in detail previously, our description here will be very brief.  Solid helium in an experimental cell is penetrated on two sides by superfluid-filled Vycor rods V1 and V2, which in turn are in contact with separate reservoirs R1 and R2 filled with superfluid.  During the experiments, a temperature gradient is present across the superfluid-filled Vycor\cite{Beamish1983,Lie-zhao1986,Adams1987} rods, which ensures that the reservoirs remain filled with superfluid, while the solid-filled cell remains at a low temperature.   For the present experiments an initial chemical potential difference, $\Delta \mu_0$, can be imposed by the creation of a temperature difference, $\Delta T = \mid T1 - T2 \mid$, between the reservoirs.  The result is a mass flux though the solid between the Vycor rods, and a change in the fountain pressure between the two reservoirs to restore equilibrium.

 To fill the cell initially,  helium gas, \emph{typically assumed} to contain $\sim$ 0.3 ppm \3he, but for this work measured to be  0.17 ppm \3he, is
 condensed through a direct-access heat-sunk capillary,
 which enters the cell at its midpoint.  To grow a solid at constant temperature from the superfluid, which is our standard technique, we begin with the pressure in the cell  just below the bulk melting pressure for \4he at the growth temperature (typically $\sim$ 300-400 mK) and then add atoms simultaneously through capillaries that enter the separate reservoirs. As with many experiments with solid \4he, we have no direct knowledge of the sample crystal quality, but presume that it has substantial sample-dependent disorder, unless annealed.

To study the effect of the \3he impurities the cell is emptied between each sample and a new concentration is introduced. To accomplish this, the cell is again filled with \4he liquid (0.17 ppm \3he).   Then a small calibrated volume at room temperature is filled with pure \3he to a known pressure. This is injected into the cell via the same direct-access capillary and is followed by additional  \4he, which also enters through the capillary, to bring the cell to the melting curve. With knowledge of the relevant volumes and pressures, a concentration of \3he is thus introduced into the cell. A solid is created (with the direct-access capillary closed) by further additions of  \4he by use of the capillary lines that enter each reservoir, the sample solidifies, and additional  \4he is added to bring the pressure of  the hcp solid to the desired range. The solid is then allowed to equilibrate, typically for several hours at $T \le 0.4$ K.

With stable solid \4he in the cell, we use heaters H1 (H2) to change the reservoir temperatures  $T1$ ($T2$) to create an initial chemical potential difference, $\Delta \mu_0$, between the reservoirs and then measure the resulting changes\cite{Ray2010b} in the reservoir pressures $P1$ and $P2$.  This allows us to determine the time dependence of the chemical potential difference, $\Delta \mu$, that drives the flux\cite{Vekhov2012}.  We take $F = d(P1-P2)/dt$ (here consistently measured at $\Delta \mu$ = 5~mJ/g) to be proportional to the flux of atoms that passes through the solid.
We report our flux values in mbar/s, where a typical value of 0.1 mbar/s corresponds to a mass flux through the cell of $\approx$ $4.8 \times 10^{-8}$ g/sec.

\begin{figure}
\resizebox{3.5 in}{!}{
\includegraphics[trim=1cm 1cm 1cm 1cm]{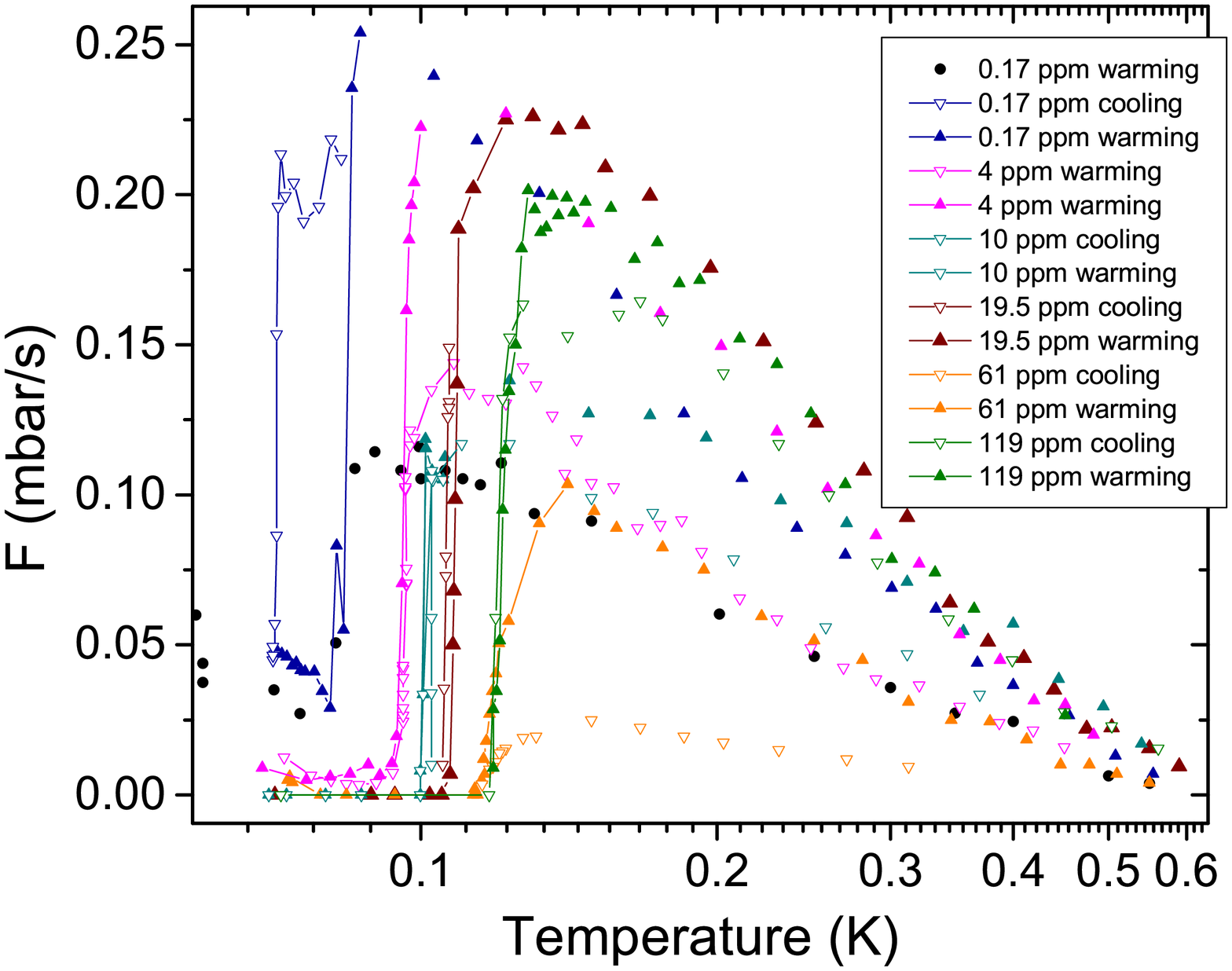}}
\caption{ \label{fig:2PRL3} (color online)
The temperature dependence of the sample-dependent flux for various concentrations.  Lines are guides to the eye. } 
\end{figure}


\begin{figure}
\resizebox{3.5 in}{!}{
\includegraphics[trim=1cm 1cm 0.2cm 1cm]{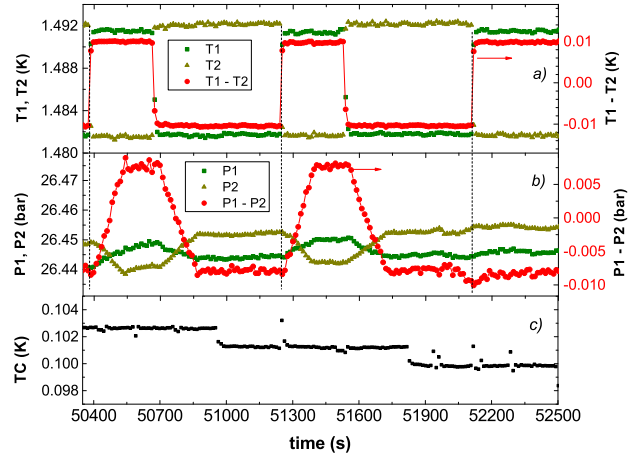}}
\caption{ \label{fig:3xPRL2} (color online)  %
An example of the sharpness of the extinction of the flux for $\chi = 10$ ppm. (a) The reservoir temperatures are reversed to initiate flow in one direction or the other; $\Delta T = T1 - T2$. (b) The resulting pressure changes allow a measurement of the $F$; $\Delta P = P1 - P2$.  (c) The cell temperature is reduced in a stepwise fashion.    }
\end{figure}

\begin{figure}
\resizebox{3.5 in}{!}{
\includegraphics[trim=1cm 1cm 0.2cm 1cm]{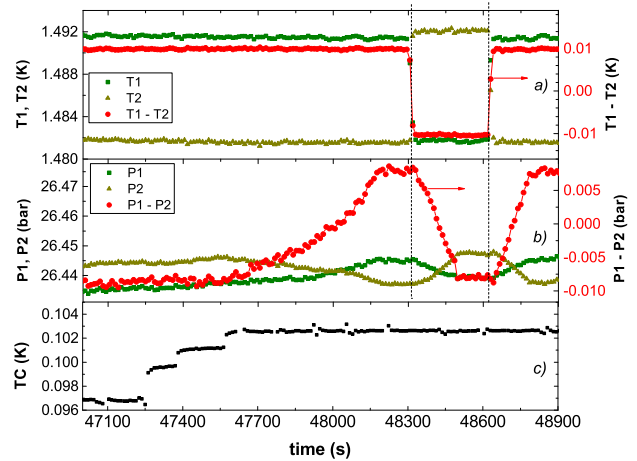}}
\caption{ \label{fig:3xPRL3} (color online)  %
Similar to the previous figure. Here in spite of an imposed $\Delta T$, no flow results until  $T$ exceeds $T_d$, after which the kinetics of the rising flux are visible. }
\end{figure}

Examples of the temperature dependence of the flux are shown in Figure~\ref{fig:2PRL3} for a number of different samples and values of $\chi$.   We document an abrupt $\chi$-dependent reduction of the flux  at a characteristic temperature, $T_d$.  Our present measurements for a $\chi$ = 0.17 ppm sample confirm the decrease in the flux in the vicinity of $75-80$ mK that was seen previously for nominal 0.3 ppm \3he\cite{Ray2010c,Ray2011a}.  We also confirm that near the foot of the drop in flux for $\chi < 5$ ppm  the flux can be rather unstable in time and after falling in a narrow temperature range can sometimes be non-zero and increase with a further decrease in temperature.
Figure~\ref{fig:3xPRL2} illustrates how sharp the flux extinction can be. In this $\chi$ = 10 ppm \3he example there is robust flux for the solid at 102.6 mK and also at 101.1 mK, but at the temperature of 99.8 mK the flux has been extinguished.  This is evident 300 seconds after the change in cell temperature to 99.8 mK, when the reversal of the applied $\Delta \mu_0$ produces no measurable flux.  An increase in the temperature of $\approx$ 1 mK results in an accelerating recovery of the flux to the previous value, Figure~\ref{fig:3xPRL3}, with a time for recovery of $\approx$~500 sec. We also note, Figure~\ref{fig:2PRL3}, that some hysteresis is present at $T_d$ and that increasing concentrations of \3he appear to cause a change in the $T_d$ flux extinction to a somewhat less precipitous behavior.


\begin{figure}
\resizebox{3.5 in}{!}{
\includegraphics[trim=1cm 1cm 1cm 1cm]{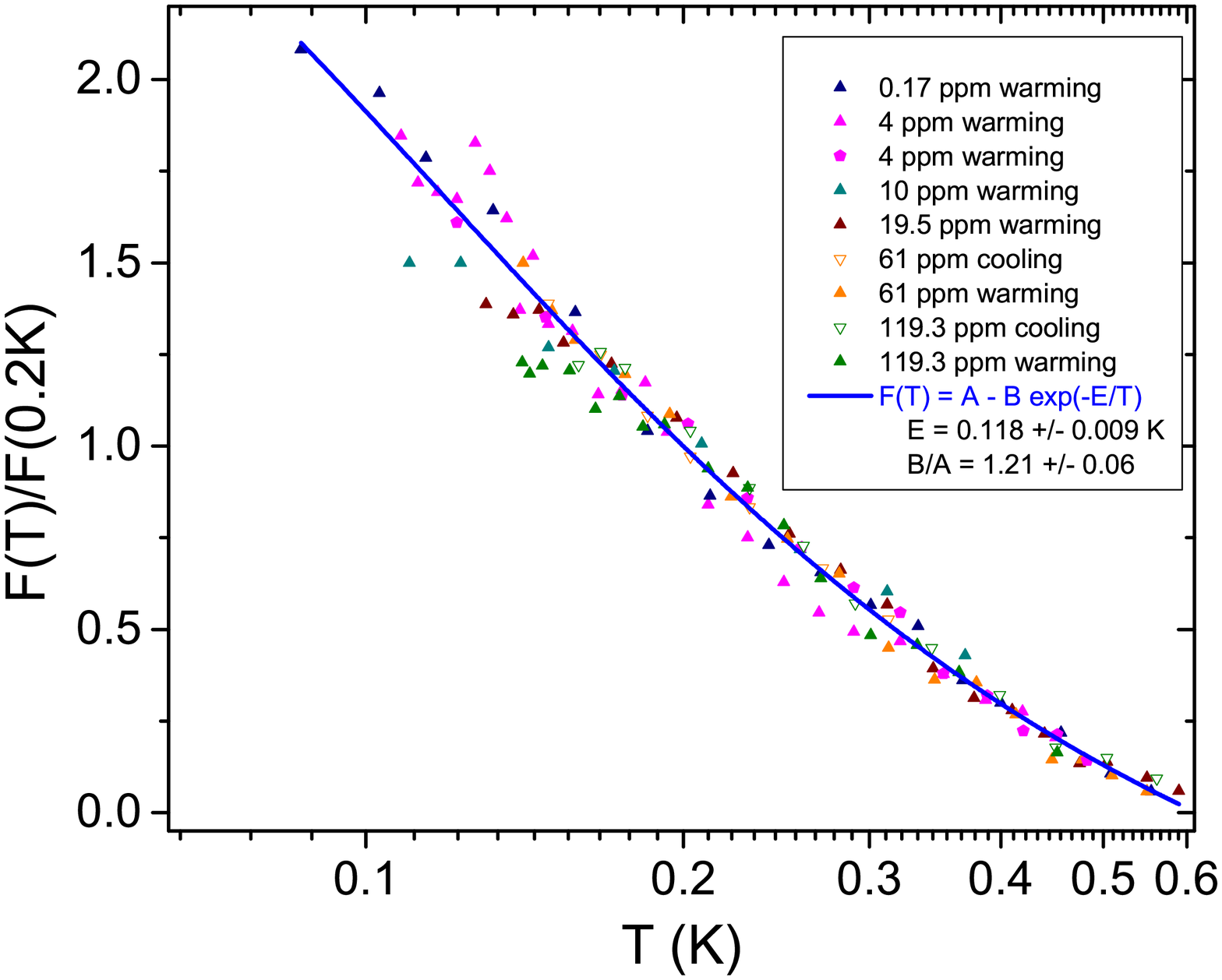}}
\caption{ \label{fig:collapse} (color online)  %
The temperature dependence of the normalized flux observed for \4he with several \3he impurity concentrations and experimental conditions for $T > T_d$. Fitted line: see text.   }
\end{figure}

The addition of \3he has no measurable effect on the non-hysteretic temperature dependence of the flux for $T > T_d$.  Different freshly-made samples typically provide different flux values.
Samples that are warmed to 500 - 650 mK or above (where the flux gets unstable or falls to zero) can show significantly lower flux when cooled -  in some cases showing no flux. Samples annealed near 1K for $\sim$~10 h show no flux when cooled and pressure gradients that existed when the sample was grown are removed.  After cooling, low (or zero) flux values can typically be increased by changing the pressure in the cell.
In all cases of non-zero flux, normalization of the data sets in at 200 mK shows that they all have the same universal-like temperature dependence as illustrated in Figure~\ref{fig:collapse}.
All of the data suggest that at higher temperature the behavior of the flux \emph{extrapolates} to zero near $T_h \approx 625$ mK, a value consistent within errors with earlier measurements\cite{Ray2010c,Ray2011a} with nominal 0.3~ppm samples.

 \begin{figure}
\resizebox{3.5 in}{!}{
\includegraphics[trim=1cm 1cm 1cm 1cm]{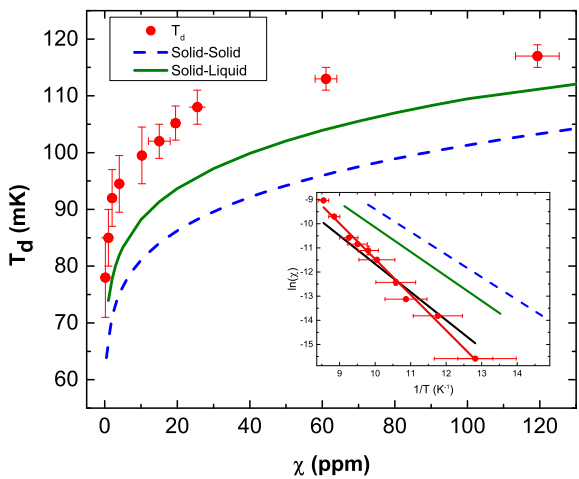}}
\caption{ \label{fig:4PRL2} (color online)  %
 Temperature of the sharp drop in $F$, $T_d$.
 Inset:  $\ln{(\chi)}$ \emph{vs.} $1/T$; see text.  }
\end{figure}

In Figure~\ref{fig:4PRL2} we illustrate the measured $T_d$ \emph{vs.} $\chi$.  The general trend of these $T_d$ \emph{vs.} $\chi$ data is \emph{reminiscent of} a phase separation curve.  With this in mind, also shown on Figure~\ref{fig:4PRL2} are the results of calculations of homogeneous phase separation.  As  a first approximation,  the coordinates of homogeneous phase separation for the  solid-solid  (\4he-rich hcp -- \3he-rich bcc; dashed line) case are obtained by use of the prescription described by Edwards and Balibar\cite{Edwards1989}:
 $T_p^s = [(0.80)(1 - 2\chi) + 0.135]/\ln(1/\chi -1)$.
In our case, this prescription needs modification since at our pressure of 25.8 bar if the \3he separates into macroscopic regions we expect that it to be liquid\cite{Ganshin1999,Huan2011,Kim2011}. We have recalculated homogeneous phase separation for the solid-liquid  (\4he-rich hcp -- \3he-rich liquid) case using the prescription of Edwards and Balibar\cite{Edwards1989} for evaluating the necessary parameters at our pressure and the corresponding $T_p^s$, Figure~\ref{fig:4PRL2} (solid line). There is limited experimental data in the literature on solid phase separation in our experimental regime\cite{Kim2011}.

The fact that an increase in concentration   shifts $T_d$ to higher values motivates a scenario for the role of the \3he in these experiments. 
We noted above that the 10 ppm sample, given a 1 mK decrease in cell temperature to 99.8 mK followed by a wait of 300 s, produced no flux following the reversal of $\Delta \mu_0$. Indeed there is evidence in the data that the transition from flow to no flow takes place within $\sim$ 150 seconds.  Given that the time required for a complete phase separation transition in solid mixture solutions is typically measured in \emph{hours}\cite{Ganshin1999,Huan2011,Kim2011}, e.g. $\sim$ 10 h, the disparity between these two times is striking.
This suggests that only a small amount of  the  \3he needs to be involved to extinguish the flux.

Since solid helium has demonstrated one-dimensional bosonic Luttinger liquid behavior\cite{Vekhov2012}, we consider the possibility  that dislocation cores and their intersections are responsible for the flux and these are blocked by the \3he.  It is predicted that the addition of \3he along a dislocation core will diminish the superfluid density there\cite{Corboz2008},
particulary where such cores intersect.  Given the number of \3he available and the likely number ($\sim 10^5$) of such structures that provide the conducting pathways between the Vycor rods\cite{Vekhov2012}, there is more than enough \3he to quickly provide for the extinction of the flux.
It is enough that a short segment or intersection along the Vycor-to-Vycor pathway that spans the cell be decorated and this should take place relatively quickly.

The inset to Figure~\ref{fig:4PRL2} shows $\ln(\chi)$ \emph{vs.} $1/T$. At small $\chi$ the bulk phase separations satisfy $\chi = \exp(-R/T)$ with $R$ approximately independent of temperature, and $R= 0.94$ K and $1.02$ K for solid-solid and solid-liquid bulk phase separation, respectively. A fit of the data (red circles, black solid line) by $\chi = \exp(-R/T)$ yields $R$= 1.17 K.  A model that includes a small number of binding sites for \3he or \4he atoms yields the functional form $\chi = \exp(a-R/T)$,
where $\exp(a)/(1+ \exp(a))$ is the minimum concentration that blocks superflow, and $R$ includes the binding energy.  With this functional form, we find a much better fit (solid red line), with $R = 1.48$ K and $a = 3.36$.  This energy value is higher than the predicted\cite{Corboz2008,Syshchenko.2010} binding energy ($\sim 0.7$ K) of single \3he atoms to dislocation cores. This supports the possibility that the flux extinction results from the \3he binding to dislocation intersections\cite{Corboz2008}, where the \3he blocks the flux.

The robust non-hysteretic and universal temperature dependence for temperatures between $T_d$ and near but below $T_h$
suggests to us the following scenario. A given sample preparation results in a given number of structures that span the sample between the Vycor rods and carry the flux.
We believe that an increase in temperature reduces the effective conductivity of the structures (which includes their connection to the superfluid-filled Vycor) that carry the flux. From this perspective conducting pathways remain robust until a high enough temperature is reached at which some (or all) of the pathways are somehow irreversibly interrupted. Indeed, as we have noted, it can be the case that an increase in the temperature of the solid well above $T_h$ leads to no flux when the cell is cooled.  This no-flux situation can be changed by an imposed change in the amount of \4he in the cell, which apparently introduces structural changes, which create new pathways for the flux.  In this scenario all of the temperature dependence well above $T_d$ is dictated by changes in conductivity along the existing pathways.

To follow this line of thought, suppose an activated process exists that degrades the flux with increasing efficiency according to $\sim \exp(-E/T)$. For example, thermally activated jogs or kinks\cite{Aleinikava.2012} (roughness) on dislocation cores would introduce disorder and phase slips would result. It is reasonable to assume that the flux might obey a form such as $F = A - B\exp(-E/T)$.  We have applied this to the data shown in Figure~\ref{fig:collapse} and we find that $B/A = 1.21 \pm 0.06$ and thus the data can be well-fit with the form $F/F(0.2K) = F_0[1 - 1.21\exp(-E/T)]$, with $E = 118 \pm 9$mK, Figure~\ref{fig:collapse}. In this scenario, when   $F = F_0^*[1 - 1.21\exp(-E/T)]$ is applied to non-normalized individual data sets, $F_0^*$ should in each case be proportional to the number of conducting pathways between the Vycor rods.


In summary, we find that the addition of \3he to concentrations above the nominal $\chi$ found naturally in well-helium serves to increase the temperature at which a sharp drop in flux through the solid-filled sample cell takes place.  We find that the temperature dependence of the flux at higher temperatures is universal
 and
the flux terminates in a narrow window near a characteristic temperature $T_h \approx$ 625 mK.
These measurements impose constraints that any explanation of the flux must satisfy and supports the possibility that the flux is carried by dislocation cores and is blocked by \3he binding.

 We thank  B.~V. Svistunov, N.~V. Prokof'ev and A.~B. Kuklov for a number of stimulating discussions, N.~S. Sullivan and D. Candela for comments on phase separation and M.~W. Ray for early work on the apparatus.  This work was supported by
   NSF DMR 12-05217,
     DMR  08-55954 and by University Research Trust Funds.

\bibliography{ref2}

\end{document}